\documentclass{elsart}

\usepackage{amssymb}
\usepackage{graphicx}
\usepackage{hyperref}

\newcommand\0{\over }  
\newcommand\5{\hat } \newcommand\6{\partial }

\newcommand{\bea}{\begin{eqnarray}}
\newcommand{\eea}{\end{eqnarray}}
\newcommand{\be}{\begin{equation}}
\newcommand{\ee}{\end{equation}}
\newcommand{\nn}{\nonumber\\ }
\newcommand{\beq}{\begin{eqnarray}}
\newcommand{\eeq}{\end{eqnarray}}
\newcommand{\Tr}{\mathrm{Tr}\,}
\newcommand{\RE}{\,\mathrm{Re}\,}
\newcommand{\IM}{\,\mathrm{Im}\,}

\begin{document}
\begin{flushright}
~\vspace{-1.25cm}\\
{\small\sf
SACLAY-T01/108\\ TUW-01-26\\
}
\end{flushright}
\vspace{0.2cm}
\begin{frontmatter}

\title{Quark number susceptibilities\\
from HTL-resummed thermodynamics
}

\author{J.-P. Blaizot},
\author{E. Iancu}
\address{Service de Physique Th\'eorique, CE Saclay,
        F-91191 Gif-sur-Yvette, France}
\author{A. Rebhan}%
\address{Institut f\"ur Theoretische Physik,
         Technische Universit\"at Wien,\\
         Wiedner Hauptstra\ss e 8-10/136,
         A-1040 Vienna, Austria
}%

\begin{abstract}
We compute analytically the diagonal quark number susceptibilities for a
quark-gluon plasma at finite temperature and zero chemical
potential, and compare with recent lattice results.
The calculation uses the approximately self-consistent
resummation of hard thermal and dense loops
that we have developed previously. For temperatures 
between 1.5 to $5T_c\,$, our results follow the
same trend as the lattice data, but exceed them in magnitude
by about $5-10\%$. We also compute the lowest order contribution,
of order $\alpha_s^3\log(1/\alpha_s)$, to the off-diagonal
susceptibility. This contribution,
which is not a part of our self-consistent
calculation, is numerically small, but not
small enough to be compatible with a recent lattice
simulation.

\end{abstract}

\end{frontmatter}

\section{Introduction}

A lot of effort is presently devoted to understanding the 
properties of hot and dense matter from Quantum Chromodynamics. This 
is motivated in part by the ongoing experimental program on 
ultrarelativistic heavy ion collisions, and also by the progress in 
lattice gauge calculations which provide so far the best theoretical 
tool at our disposal to calculate from first principles the properties 
of the quark-gluon plasma. Recently however, it has been shown that 
results of such calculations could be remarkably well reproduced by weak 
coupling techniques when the temperature is larger than 2 to 3 times
the transition temperature
\cite{Blaizot:1999ip,Blaizot:1999ap,Blaizot:2000fc}. 
The purpose of this paper is to apply these techniques 
to the calculation of quark-number susceptibilities which have 
recently received considerable attention.

These quantities are interesting in several respects. First of all, 
they are to date
about the only quantities that can be calculated on the 
lattice and provide information about finite density
\cite{Gottlieb:1987ac,Gavai:1989ce,Gottlieb:1997ae,Bernard:1996zw,Gavai:2001fr,Gavai:2001ie}. (Recall that 
lattice calculations are still limited to zero chemical potential;
susceptibilities involve derivatives of the thermodynamic functions 
with respect to $\mu$, and their limit as $\mu\to 0$
can be computed on the lattice.)
Susceptibilities have also been discussed lately 
in the context of heavy ion collisions, as they can be related to 
measurable fluctuations in conserved quantities
\cite{Asakawa:2000wh,Jeon:2000wg,Koch:2001zn}. 
However, the main question addressed here is a theoretical one,
namely, whether the recent lattice results 
in Refs. \cite{Gavai:2001fr,Gavai:2001ie}
can be explained within resummed perturbation theory, that is,
without invoking genuine non-perturbative contributions.

The lattice results \cite{Gavai:2001fr,Gavai:2001ie}
for the diagonal susceptibility  $\chi$ 
(cf. eqs.~(\ref{def-chi})--(\ref{offdiag}) below)
 at temperatures between $1.5$ and $5T_c$
show a slow approach of the ideal-gas result 
from below, with deviations of about $15\%$. But 
the weak coupling expansion of $\chi$ completely fails to reproduce
this behaviour. In massless QCD at $\mu=0$, this expansion
is presently known to order $g^4 \log(1/g)$ 
\cite{Toimela:1985xy,Kap:FTFT} :
\begin{eqnarray}\label{cc0pt}
  {\chi \0 \chi_0}&=&1-{1\02}{3\0N}{N_g\08}\left(g\0\pi\right)^2+
{3\0N}{N_g\08}\sqrt{{N\03}+{N_f\0 6}}\left(g\0\pi\right)^3 \nn
&&-{3\04}{N_g\08}\left(g\0\pi\right)^4\log{1\0g}+ {\mathcal O}(g^4)
\end{eqnarray}
(with $\chi_0 =NT^2/3$ the ideal gas value and $N_g=N^2-1$).
Leaving aside the still undetermined $g^4$ contribution, one finds that
the perturbative results lie {\it above} the ideal gas values
for all temperatures of interest, and decrease with increasing $T$. 

We may relate this failure to that
encountered in the perturbative calculation of the
pressure \cite{Braaten:1996ju}.
In both cases, the difficulty of perturbation theory
has its origin in 
collective 
phenomena which develop at the scale $gT$, and which in a
strict perturbative expansion provide large contributions 
starting at order $g^3$.
To cope with this, various resummation schemes have been proposed
\cite{Blaizot:1999ip,Blaizot:1999ap,Blaizot:2000fc,Andersen:1999fw,%
Baier:1999db,Peshier:2000hx}.
Here, we shall use the one developed in  
Refs. \cite{Blaizot:1999ip,Blaizot:1999ap,Blaizot:2000fc},
which focuses on the physical picture of the quark-gluon plasma
as a gas of quasiparticles with properties determined
by the ``hard thermal loops'' (HTL) \cite{Blaizot:2001nr,Rebhan:2001wt}.
This approach has proven to be successful
in describing the lattice data for the
thermodynamics of QCD down to temperatures as low as $2.5T_c$.

First lattice
measurements of the {\it off-diagonal} susceptibility $\chi_{ud}$
have also been reported in Ref.~\cite{Gavai:2001ie}.
This quantity vanishes for the ideal gas, so it probes
directly the interactions in the system.
We show that, when $\mu=0$, it is of order $g^6\log(1/g)$,
and we compute this lowest-order contribution in both QCD and QED.

\section{Diagonal susceptibilities}
\label{diag}

Quark number susceptibilities are generally defined as:
\be\label{def-chi}
\chi_{ij}\equiv {\6 {\mathcal N}_i \0 \6 \mu_j}
 = {\6^2 {P} \0 \6 \mu_i\6 \mu_j}=
\chi_{ji}
\ee
where $i,\,j$ are flavor indices, 
${\mathcal N}_i $ is the quark number density, and $P$ is the pressure.
With all quarks massless and $\mu_i=0$ (as appropriate 
for comparison with the lattice results), all diagonal and
all off-diagonal elements become equal, and we write 
\bea\label{offdiag}
\chi_{ij}\Big|_{\mu=0} \equiv \chi & \quad \mbox{for $i=j$}\,, \qquad
\chi_{ij}\Big|_{\mu=0} \equiv \tilde\chi & \quad \mbox{for $i\not=j$}.
\eea

We shall evaluate the diagonal susceptibility $\chi$
within the resummation scheme developed in Refs.
\cite{Blaizot:1999ip,Blaizot:1999ap,Blaizot:2000fc}.
This is based on the following expression for the 
fermion number density ${\mathcal N}$ in
terms of the dressed fermion propagators $\Delta_\pm$
(cf. eqs. (4.12) and (4.19) of Ref.~\cite{Blaizot:2000fc}):
\beq\label{NF}
{\mathcal N}&=& -4N\int\!\!{d^4k\0(2\pi)^4}{\6f(\omega)\0\6\mu}\,
\Bigl\{\IM\log\Delta_+^{-1}\,+\,\IM\log(-\Delta_-^{-1})\,+\nn
&{}&\qquad\qquad\qquad \,-\,\IM\Sigma_+\RE\Delta_+\,+\,
\IM\Sigma_-\RE\Delta_-\Big\},\eeq
where $\Delta_\pm^{-1} \equiv - [\omega\mp(k+\Sigma_\pm)]$, 
$\Sigma_\pm$ are the corresponding self-energies, and the plus 
(minus) subscript applies to fermions whose chirality is equal 
(opposite) to their  helicity. The fermion self-energies
and propagators are diagonal in flavor indices, and
eq.~(\ref{NF}) applies to 
each quark flavor $i$ separately, 
but flavor indices are kept implicit.

As in Refs. \cite{Blaizot:1999ip,Blaizot:1999ap,Blaizot:2000fc},
we shall consider two successive approximations to the
self-energies $\Sigma_\pm$.
The first is the HTL approximation where \cite{Blaizot:2001nr}:
 \beq\label{SIGHTL}
\hat\Sigma_\pm(\omega,k)\,=\,{\hat M^2\0k}\,\left(1\,-\,
\frac{\omega\mp k}{2k}\,\log\,\frac{\omega + k}{\omega - k}
\right),\eeq
and $\hat M^2$ is the plasma frequency for fermions,
i.e., the frequency of long-wavelength ($k\to 0$) fermionic
excitations ($C_f=(N^2-1)/2N$):
\be\label{MF}
\hat M^2=\frac{g^2 C_f}{4\pi^2}\int_{0}^\infty 
{\rm d}k \,k\Bigl(2n(k)+f_+(k) + f_-(k)\Bigr) =
{g^2 C_f\08}\left(T^2+{\mu^2\0\pi^2}\right).\ee
In this approximation, there is no mixing between quarks of 
different flavors, so the corresponding susceptibilities are
diagonal even for $\mu\ne 0$.

The resulting expression of the number density, denoted by
 ${\mathcal N}_{HTL}$, is the sum of two contributions:
${\mathcal N}_{HTL}
= {\mathcal N}_{HTL}^{\rm QP}+{\mathcal N}_{HTL}^{\rm LD}$,
where ${\mathcal N}_{HTL}^{\rm QP}$ is the contribution
of the quasiparticle poles\footnote{Charge
conjugation ($\Sigma_+(\omega,k)=\Sigma_-(-\omega,k)$) exchanges the poles of 
$\Delta_+$ and $\Delta_-\,$: $\Delta_+$ has 
{two} poles, one at positive $\omega$, with energy
$\omega_+(k)$,  and another one at negative $\omega$, with energy
$-\omega_-(k)$; these go over to $\pm \hat M$ as $k\to 0$.
Correspondingly, $\Delta_-$ has poles at
$\omega_-$ and $-\omega_+$. See, e.g., Ref. \cite{Blaizot:2001nr}.}
$\omega_\pm=\pm[k+\5\Sigma_\pm(\omega_\pm,k)]$, and
${\mathcal N}_{HTL}^{\rm LD}$ is that of
the Landau damping cuts at $-k<\omega<k$. We have:
\bea\label{NfQP}
{\mathcal N}_{HTL}^{\rm QP}&=&N\int\limits_0^\infty{ k^2 dk \0 \pi^2}
{\6\0\6\mu} \Bigl\{ T\log(1+e^{-[\omega_+(k)-\mu]/T}) \nn
&+& T\log{1+e^{-[\omega_-(k)-\mu]/T} \0 1+e^{-(k-\mu)/T}} 
+(\mu\to-\mu) \Bigr\},
\eea
where the $\mu$ derivative is to be applied to 
the explicit $\mu$ dependence only, and not to that
implicit in the dispersion laws 
$\omega_+(k)$ and $\omega_-(k)$ of the quasiparticles.
Similarly,
\bea\label{NfLD}
&&{\mathcal N}_{HTL}^{\rm LD}=-N\int\limits_0^\infty{ k^2 dk \0 \pi^3}
\int\limits_0^k \! d\omega \left[{\6f_+(\omega)\0\6\mu}+
{\6f_-(\omega)\0\6\mu}\right]  \nn
&&\quad\times \Bigl\{ \arg[k-\omega+\hat\Sigma_+(\omega,k)] 
-\IM\hat\Sigma_+(\omega,k) \RE[k-\omega+\hat\Sigma_+(\omega,k)]^{-1} 
\nn&&\quad+\arg[k+\omega+\hat\Sigma_-(\omega,k)]
-\IM\hat\Sigma_-(\omega,k) 
\RE[k+\omega+\hat\Sigma_-(\omega,k)]^{-1}
\Bigr\}.
\eea

\begin{figure}
\includegraphics[bb=70 405 285 500]{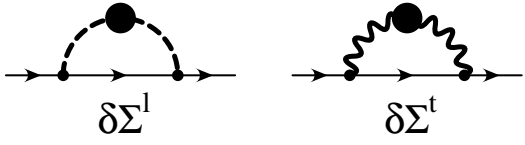}
\caption{NLO contributions to $\delta\Sigma$ at hard momentum. Thick
dashed and wiggly lines with a blob represent HTL-resummed longitudinal
and transverse gauge boson propagators, respectively.
\label{figdSigma}
}
\end{figure}

The HTL approximation  \cite{Blaizot:2001nr} contains the perturbative
contributions of order $g^2$. This comes exclusively from
the hard ($k\sim T$), ``normal'', branch
$\omega_+$ and its
asymptotic thermal mass $M_\infty^2\equiv 2k \hat\Sigma_+(\omega=k) =
2 \hat M^2$ :
\be
{\mathcal N}^{(2)}\,=\,-\,{N\02\pi^2}\,\mu M_{\infty}^2\,.
\ee
However, there is no $g^3$ contribution in ${\mathcal N}_{HTL}$.
Such a contribution, denoted as ${\mathcal N}^{(3)}$, comes entirely
from the next-to-leading
(NLO) correction $\delta M_{\infty}^2(k)\equiv 2k\RE\delta\Sigma_+(\omega=k)$
to the asymptotic mass of the hard fermion 
\cite{Blaizot:1999ap,Blaizot:2000fc}:
\be\label{NLA}
{\mathcal N}^{(3)}\,=\,-4N
 \int\!\!{d^3k\0(2\pi)^32k}\ 
{\6f_+(k)\0\6\mu}\RE 2k\delta\Sigma_{+}(\omega=k),\ee
where the NLO self-energy $\delta\Sigma_{+}$ 
is given by the diagrams in Fig.~\ref{figdSigma}.

In contrast to the lowest order asymptotic mass $M_{\infty}^2$,
the correction $\delta M_{\infty}^2(k)$
is a nontrivial function of the momentum \cite{Blaizot:2000fc} which 
can be evaluated only numerically. However, this function contributes
to eq.~(\ref{NLA})
only in an averaged form \cite{Blaizot:1999ap,Blaizot:2000fc} :
\be\label{deltaMas}
\bar\delta M_\infty^2={\int dk\,k\,[\6 f_+(k)/\6\mu] \RE 
2k \delta\Sigma_+(\omega=k) 
\0 \int dk\,k\,\6 f_+(k)/\6\mu}\,=\,-\,{1\02\pi}\,g^2C_fT\hat m_D\,,
\ee
where $\5m_D$, the Debye mass, is
\be\label{mD}
\5m_D^2=(2N+N_f)\frac{g^2T^2}{6}+\frac{g^2}{2\pi^2}\sum_j{\mu_j^2}\,.\ee
Thus, at a strictly perturbative level, it would be possible
to reproduce the perturbative result for ${\mathcal N}$ to
order $g^3$ by replacing $2 \hat M^2\equiv
M_{\infty}^2\rightarrow M_{\infty}^2 + \bar\delta M_\infty^2$ 
in eqs.~(\ref{NfQP},\ref{NfLD}) 
for the HTL approximation to ${\mathcal N}$.

However, the correction  $\bar\delta M_\infty^2$ is negative, and
for $g \gtrsim 1$ it is of the same
order of magnitude or larger than the lowest-order asymptotic
mass, apparently leading to a tachyonic thermal
mass. As we have argued previously 
\cite{Blaizot:1999ip,Blaizot:1999ap,Blaizot:2000fc},
this problem is not specific to QCD, but can be studied already in
simple scalar $g^2\varphi^4$ theory.
There the perturbative thermal mass to NLO is
$m^2=g^2T^2(1-3g/\pi)$. The corresponding one-loop
gap equation, on the other hand, gives a monotonic function
of $g$, which is well approximated by the
quadratic equation \cite{Blaizot:2000fc} 
$m^2=g^2T^2-{3}mT/\pi$. Also 
a simple Pad\'e resummation \cite{Blaizot:1999ip} $m^2=g^2T^2/(1+3g/\pi)$
gives reasonable approximations
even for $g\gg 1$. In the following, we shall consider
both prescriptions for including
(\ref{deltaMas}) into the asymptotic
thermal mass, referring to them by ``NLQ'' and ``NLP'',
respectively.

\section{Numerical evaluation}

\begin{figure}
\includegraphics[bb=80 160 500 500,width=9cm]{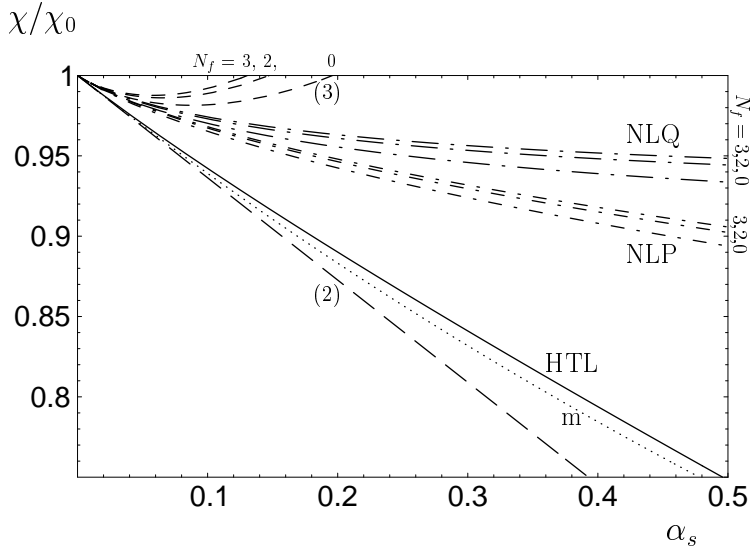}
\caption{Quark number susceptibility normalized to its
free-field value in SU(3) with $N_f=0,2,3$ quark flavors
as a function of $\alpha_s$.
Dashed lines give the perturbative results through order
$\alpha_s$ (marked ``(2)'') and order $\alpha_s^{3/2}$
(marked ``(3)''). The full line gives our result using
HTL propagators, the dotted one a simpler quasiparticle model
with momentum-independent mass $M=M_\infty$ (both of which
are $N_f$ independent).
Including next-to-leading order corrections to the
asymptotic fermion mass through a quadratic
gap equation gives the lines marked ``NLQ''; using
instead a simple Pad\'e approximant gives the lines
marked ``NLP''.\break
\label{figccal}}
\end{figure}

The results of a numerical evaluation of $\chi/\chi_0$ are given in 
Fig.~\ref{figccal} as a function of $\alpha_s$.

The HTL approximation gives results
which are above those of first-order perturbation theory (i.e.,
order $g^2$).
Since the former does not include anything of the plasmon
effect $\propto g^3$, the visible deviation is to be attributed
to higher order contributions. A numerical analysis reveals that
most of the enhancement is due to terms of order $g^4$, which
also involve a logarithm :
\be\label{g4HTL}
\chi_{HTL}^{(4)}|_{\mu=0}= N \left(
0.0431\ldots \times\log {T\0 \hat M} + 0.0028\ldots \right){\hat M^4\0T^2}.
\ee
The coefficient of the log is by a factor of $\approx -0.52 (N^2-1)/N^2$ 
different from that of the perturbative result (\ref{cc0pt}). The
correct coefficient will
be restored by $O(g^4 \log(1/g) T^2)$ corrections
to $M_\infty^2$ (not considered here)\footnote{The constant behind the
logarithm (which is still unknown in
perturbation theory) receives three-loop contributions which are
beyond the $\Phi$-derivable two-loop approximation underlying
the density expression (\ref{NF}).}.

It is instructive at this stage to compare with the susceptibility
of an ideal gas of massive fermions with mass equal to the asymptotic 
HTL mass, and which therefore contains the correct contribution of
order $g^2$. This reads (with $\omega_k=\sqrt{k^2+2\hat M^2}\,$) :
\be\label{ideal}
\chi_{\mathrm{m}}|_{\mu=0}=
{2\0T}\int_0^\infty {dk\,k^2\0\pi^2} {e^{\omega_k/T}
\0(e^{\omega_k/T}+1)^2} = 2 {\6\0\6\log T} 
\int_0^\infty {dk\,k^2\0\pi^2 \omega_k}{1\0e^{\omega_k/T}+1}\,.
\ee 
In contrast
to  eq.~(\ref{g4HTL}), this does not involve any
logarithmic term at order $g^4$
\be
\chi_{\mathrm{m}}^{(4)}|_{\mu=0}= N {7 \zeta(3) \0 4 \pi^4}{\hat M^4\0T^2}
\approx 0.0216 N{\hat M^4\0T^2}.
\ee
Numerically, however, (\ref{ideal}) happens to be 
rather close to the HTL expression, 
as can be seen from the dotted line in Fig.~\ref{figccal}.

At any rate, these order $g^4$ effects in either (\ref{g4HTL}) or
(\ref{ideal}) are quite small compared to the more decisive
order $g^3$-contribution. As we have seen,
in the self-consistent density the effect of order $g^3$
comes exclusively from the NLO correction to the asymptotic
thermal mass. This introduces a (weak) 
dependence upon $N_f$, via the
the Debye mass (\ref{mD}). 
As an estimate of this  effect, we
include it in the averaged form (\ref{deltaMas}),
for simplicity by a rescaling of $\hat M$ for all momenta.
In order to get an idea of the theoretical uncertainties, we
do so 
alternatively through a quadratic gap equation (NLQ) or
through a (2,1)-Pad\'e approximant (NLP). 
The corresponding numerical results for $N_f=0,2,3$
are shown in Fig.~\ref{figccal} by the
various dash-dotted lines, with the formal limit $N_f=0$ 
corresponding to the quenched approximation of
lattice gauge theory. As manifest on this figure, 
the inclusion of the order--$g^3$ contribution in our
self-consistent calculation has a significant effect, 
although not as dramatic as in conventional perturbation theory.

\begin{figure}
\includegraphics[bb=30 200 540 520,width=10cm]{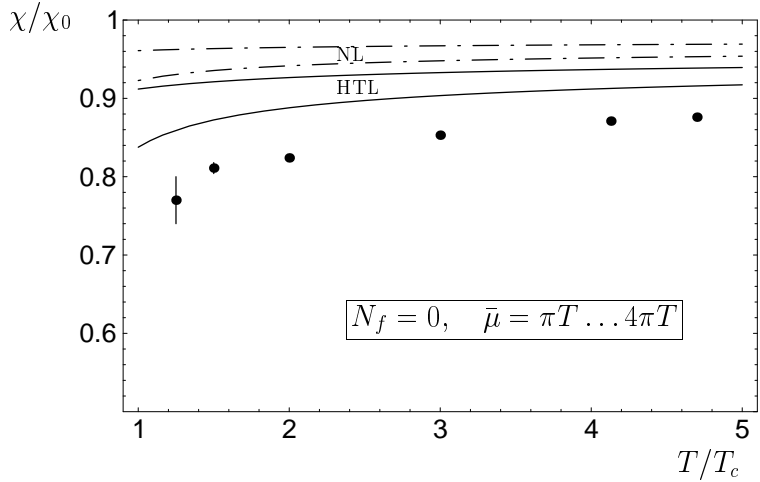}
\caption{\label{figchiq}
Comparison of our results for $\chi/\chi_0$ in massless QCD
in the formal limit $N_f=0$ 
(using $T_c/\Lambda_{\overline{\mbox{MS}}}=1.15$ \cite{Gupta:2000hr})
with the lattice results of Ref.~\cite{Gavai:2001fr} 
for quenched QCD obtained
with quark mass $m=0.12 T_c$ on a lattice with $N_t=4$ (no
continuum extrapolation). [The two rightmost
lattice data are unpublished, private communication by S. Gupta.] }
\end{figure}

\begin{figure}
\includegraphics[bb=30 200 540 520,width=10cm]{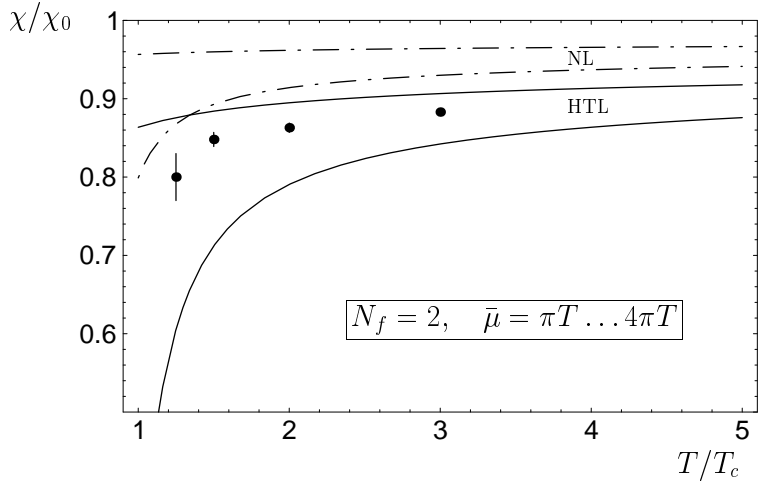}
\caption{\label{figchiNf2}
Comparison of our results for $\chi/\chi_0$ in massless $N_f=2$ QCD
(using $T_c/\Lambda_{\overline{\mbox{MS}}}=0.49$ \cite{Gupta:2000hr})
with the lattice results of Ref.~\cite{Gavai:2001ie} obtained
with quark mass $m=0.1 T_c$ on a lattice with $N_t=4$ (no
continuum extrapolation).}
\end{figure}

In Figs.~\ref{figchiq} and \ref{figchiNf2} these numerical
results are translated into plots of $\chi/\chi_0$ as a
function of $T/T_c$ using the recent determination
of $T_c/\Lambda_{\overline{\mbox{MS}}}$  of Ref.~\cite{Gupta:2000hr}
(which is found to differ significantly
for quenched QCD and $N_f=2$), together with a standard
two-loop running coupling $\alpha_s(\bar\mu)$. We vary the renormalization
scale $\bar\mu$
around $\bar\mu=2\pi T$ by a factor of 2. For an error estimate
of the NL approximations, we in addition combine the (overlapping)
results for NLP and NLQ. 

A completion of the $g^4\log(1/g)$ contributions, which 
is in principle possible
within our approach and is left for future improvements,
should decrease the NL results somewhat
and presumably bring it nearer to the HTL result.

Also given in
Figs.~\ref{figchiq} and \ref{figchiNf2} are the recent lattice results
of Refs.~\cite{Gavai:2001fr} and \cite{Gavai:2001ie}, respectively.
These results involve finite but small quark masses, and, perhaps
more importantly, are obtained for a lattice with only 4 sites in
the temporal direction, and are still waiting for a proper
continuum extrapolation. Our results follow the same general trend
as the lattice data (they slowly increase towards the ideal gas value),
but exceed the latter by some +10\%. (Remarkably, this discrepancy
is less pronounced for the physical case of dynamical fermions.)
By contrast, the perturbative result to order
$g^3$, eq.~(\ref{cc0pt}), decreases as a function of $T/T_c$
in the range studied here and, actually, up to temperatures as high
as $T\sim 100 T_c$ for $N_f=0$, and even higher for $N_f=2$.

\section{Off-diagonal susceptibilities}
\label{off-diag}

The systematics of the 
diagrammatic contributions to susceptibilities (in particular,
the off-diagonal ones) can be clarified by referring to the symmetry under
charge conjugation, or $C$--parity. Chemical potentials couple
to the fermion fields in the same way as the $A^0$ component
of an Abelian gauge field. Thus, when expanding  a quark loop in
powers of $\mu$, one may attribute to each factor of $\mu$ the 
$C$--parity of the photon field, i.e., $C=-1$.
Gluons attached to a quark loop in a colour symmetric state
behave under permutations in the same way as photon insertions,
and thus can be ascribed $C=-1$ as well.
$C$--parity conservation forbids
a photon to decay into two gluons: a colourless 2-gluon state
is necessarily colour symmetric, and therefore $C$--even. 
However, a photon can decay into three gluons which are in a colour
symmetric state, or in two gluons 
and an arbitrary odd number of photons, etc.
In terms of chemical potentials, this means that a quark loop with two
gluon external lines is necessarily even in $\mu$, while a 
quark loop with three gluon legs may generate also a term 
linear in $\mu$, which is then symmetric in the colour indices.

The first perturbative contributions
to the nondiagonal susceptibility $\tilde\chi$ require two
fermion loops connected by gluon lines. The diagram
with just one gluon exchange vanishes by colour neutrality.
The one with two gluon exchange is non-zero,
but because the fermion loops are then even functions
of $\mu$, it contributes to $\tilde\chi$ only when $\mu\not=0$,
starting at order $g^3$. One easily finds
\be
\chi_{ij}={g^4 (N^2-1)T\mu_i \mu_j \0 16 \pi^5 m_D} \qquad \mbox{for
$i\not=j$}\,.
\ee
In fact, this is the same as $\chi_{ij}=\6 {\mathcal N}^{(3)}_i/
\6 \mu_j$ ($i\not=j$) with ${\mathcal N}^{(3)}$ given in
eq.~(\ref{NLA}). From the perspective of eq.~(\ref{NLA}),
the mixing between different flavors is induced by the
resummation of quark loops along the soft, internal, gluon lines
in the diagrams in Fig.~\ref{figdSigma}.

\begin{figure}[ht]
\bigskip
\includegraphics[width=5cm]{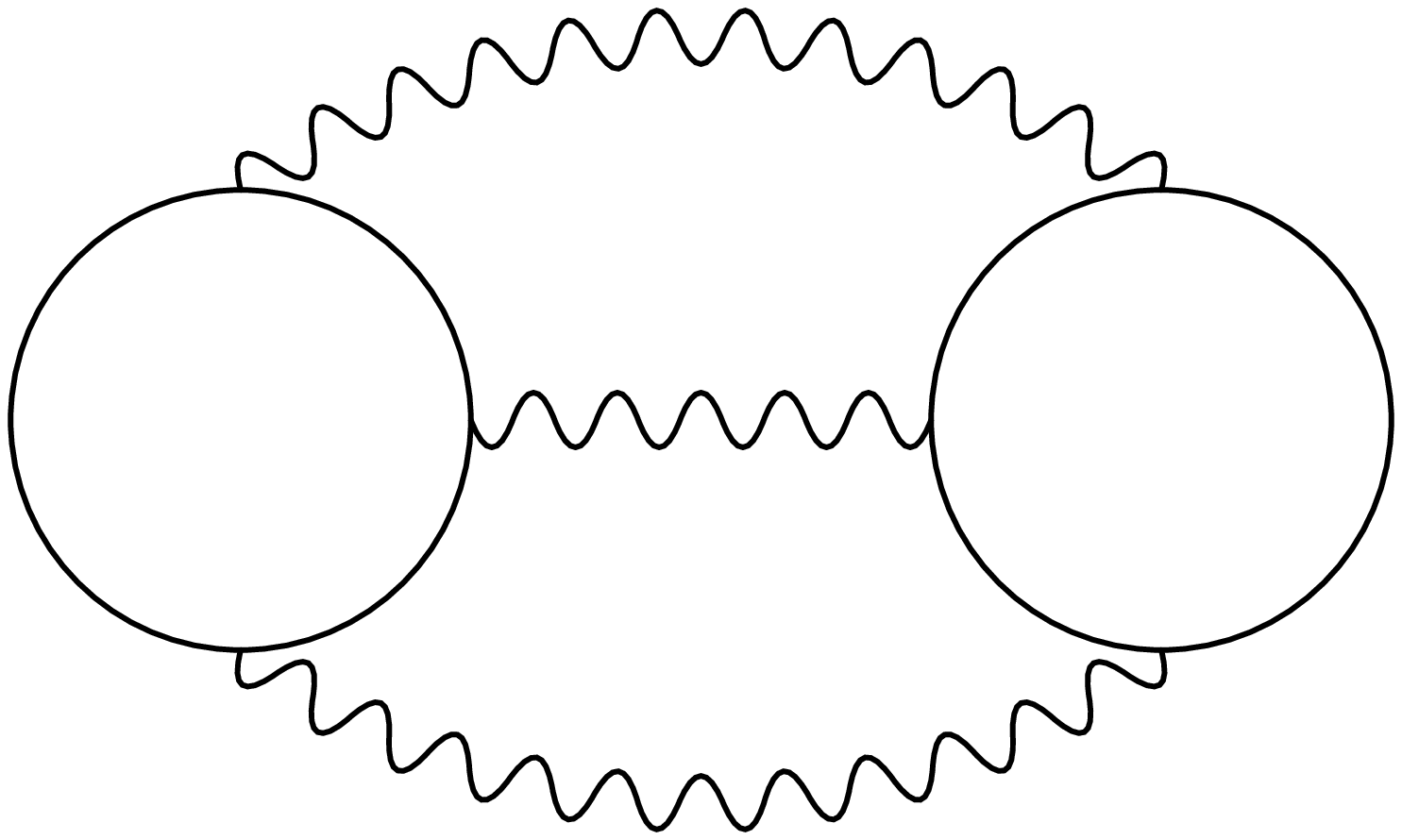}\qquad\qquad\includegraphics[width=4.6cm]{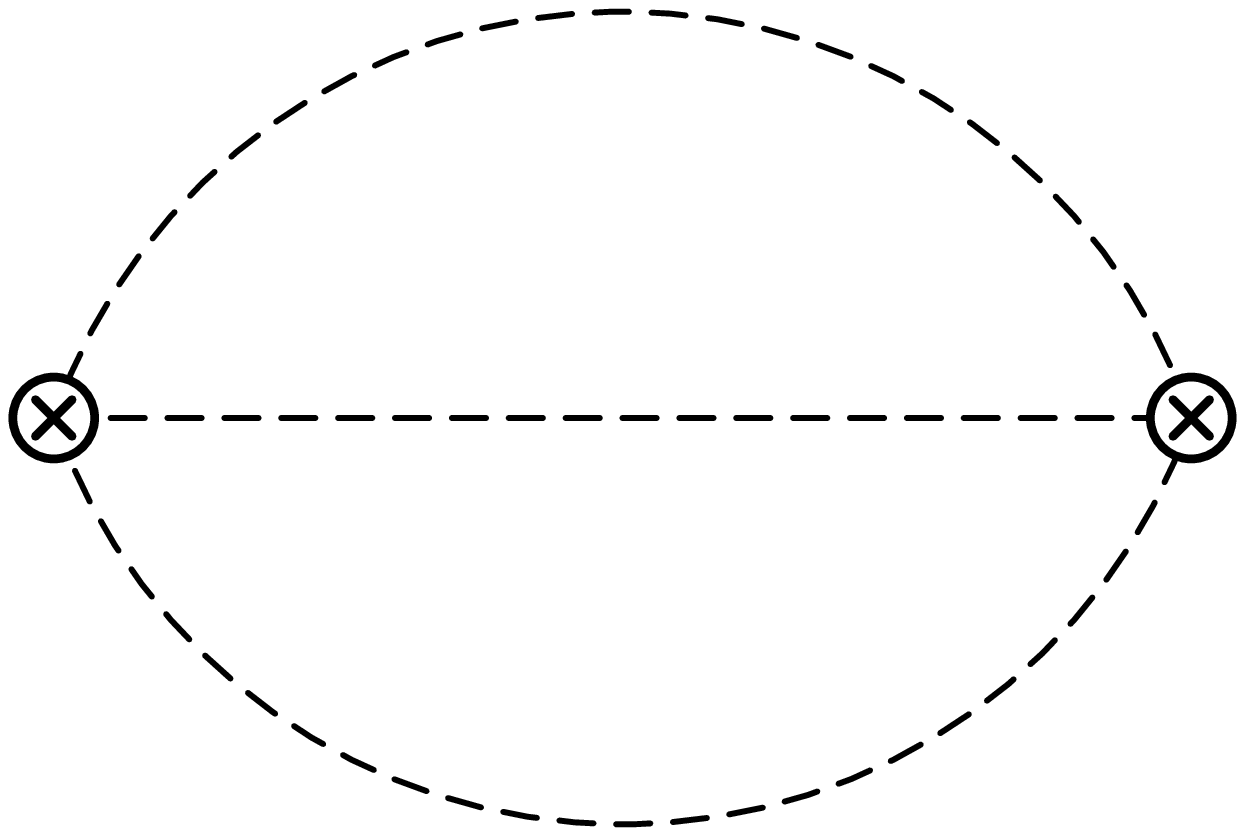}
\centerline{\hfil (a) \hfil\hfil\hfil\hfil\hfil\hfil (b) \hfil\hfil\hfil\hfil}
\caption{(a) Lowest-order diagram in the thermodynamic potential
that contributes to off-diagonal susceptibilities
$\6{\mathcal N}_i/\6\mu_j$ at $\mu=0$. 
(b) Corresponding diagram
in the effective theory for the electrostatic modes.
\label{bugblatter}
}
\end{figure}

However, when all chemical potentials vanish, the
lowest-order diagram contributing to
$\tilde\chi$ is the 
``\href{http://www.igd.fhg.de/~jasnoch/hitchhiker/Guide/g20.html}{%
bugblatter}'' \cite{Ada:HHGG} diagram
shown in Fig.~\ref{bugblatter}a. 

This diagram is superficially of order $g^6$, but when
calculated with bare gluon propagators, it 
 develops a logarithmic infrared divergence in the
electrostatic sector, because, for static external gluons,
the quark loop induces an effective local vertex corresponding to
\be\label{eff3m}
{g^3\03!}\Tr A_0^3 \sum_i {\6^3\0\6\mu^3_i}P_0^f(m_i;\mu_i,T)
\ee
where $P_0^f$ is the ideal gas pressure of a fermion
with mass $m_i$ and chemical potential $\mu_i$. 
For massless fermions this reduces to
\cite{KorthalsAltes:1999cp,Hart:2000ha,Bodeker:2001fs} :
\be\label{eff3}
{g^3\03\pi^2} \Tr A_0^3\sum_i{\mu_i} \,=\,
{g^3\012\pi^2} d^{abc}A^a_0 A^b_0 A^c_0\sum_i{\mu_i} 
\ee
($A_\mu=A_\mu^a t^a$, $\Tr t^a t^b=\delta^{ab}/2$).
We expect Debye screening to
cut off this divergence at the scale $\5m_D\sim gT$, with the
upper scale in the logarithm, of order $T$, set by the thermal
distribution. This gives a contribution to 
$\tilde\chi$ of order $g^6\log(1/g)$, 
which is the leading order effect for $g$  small enough.
We now compute its coefficient.

In the imaginary
time formalism, the infrared divergence is isolated in
the static Matsubara sector. The original diagram in 
Fig.~\ref{bugblatter}a is then identified with the two-loop
diagram in Fig. \ref{bugblatter}b. Formally, this is the
second order perturbative correction $f_2$ to the free energy 
$f=-\log Z$ of a 3-dimensional scalar field with
effective action
\be\label{SEFF}
S_E\,=\,\int d^3 x\, \frac{1}{2}\,A^a_0(-\nabla^2 +\5m_D^2)A^a_0\,
+\,i\sum_j{\mu_j} \,{g^3\sqrt{T}\012\pi^2}\,d^{abc}A^a_0 A^b_0 A^c_0,\ee
where $A^a_0({\bf x})=\sqrt{T}A^a_0(\omega_n=0,{\bf x})$,
$\5m_D$ is the Debye mass (\ref{mD}),
and the interaction term is now purely imaginary, as a consequence
of the  continuation $A_0^M\to iA^E_0$ to imaginary time.
Denoting by $S_I$ the interaction term in eq.~(\ref{SEFF}),
we have $f_2=-\langle S_I^2/2 \rangle_0$, which is {\it positive},
a consequence of the interaction term being purely imaginary.
A direct calculation yields:
\bea\label{p0}
\frac{f_2}{V}=3d^{abc}d^{abc}\,\left(
\sum_j{\mu_j} \,{g^3\sqrt{T}\012\pi^2}\right)^2
\int{d^3k\; d^3q\0(2\pi)^6} \, D_{00}(k)
D_{00}(q) D_{00}(|\mathbf k+\mathbf q|),
\eea
where $D_{00}=1/(k^2+\5m_D^2)$
and $d^{abc}d^{abc}=(N^2-1)(N^2-4)/N$.
The above integral has a spurious ultraviolet divergence, which comes
from the restriction to the static Matsubara modes, and, in the absence
of the Debye mass, it would be also divergent in the infrared.
This yields
\bea
&&\int_0^\Lambda\! {d^3k\; d^3q\0(2\pi)^6}\,{1\0[k^2+\5m_D^2] [q^2+\5m_D^2] 
[(\mathbf k+\mathbf q)^2+\5m_D^2]} \nn
&&\simeq {1\016\pi^2} \log{\Lambda\0\5m_D}\,\simeq \,{1\016\pi^2} \log{1\0g}\;,
\eea
where the upper cut-off $\Lambda$ eventually gets replaced by $T$
upon inclusion of the nonstatic Matsubara modes.

Putting everything together and returning to the original 4-dimensional 
gauge theory at finite temperature, the above estimate for
$f_2$ translates into the following, { negative},
contribution to the pressure:
\be\label{DeltaP}
\Delta P\,=\,-\frac{T}{V}\,f_2\,=\,- \,{(N^2-1)(N^2-4)\0 768 N} \,
T^2 \biggl(\sum_j \mu_j\biggr)^2 \left({g\0\pi}\right)^6  \log{1\0g}\,.
\ee


From eq.~(\ref{DeltaP}) we finally deduce the 
leading-order term for nondiagonal $\chi\,$:
\be\label{tcc0}
{\tilde \chi\0\chi_0}\simeq-
{(N^2-1)(N^2-4)\0 128 N^2}\left({g\0\pi}\right)^6 \log{1\0g}.
\ee
This logarithmically enhanced contribution
vanishes in SU($2$) gauge theories---though not in QED. The 
(ultrarelativistic) QED result is obtained by replacing 
$d^{abc}d^{abc}\to 16$ in eq.~(\ref{p0}) (cf.
eq.~(\ref{eff3})), and $\chi_0\to T^2/3$,
yielding
\be
\tilde \chi/\chi_0\Big|_{\mathrm{QED}} \simeq- {e^6\08 \pi^6} \log(1/e) 
\simeq- 4(\alpha/\pi)^3 \log(1/\alpha).
\ee

In (massless) QCD, the leading-order
contribution to $\tilde\chi$ is
\be\label{QCDtildechi}
{\tilde \chi\0\chi_0}\Big|_{N=3}
\simeq-{5\0144}\,\left({g\0\pi}\right)^6 \log{1\0g} 
\simeq-{10\09\pi^3}\,\alpha_s^3 \log(1/\alpha_s).
\ee
While in QED it is plausible that the leading-log term with its
negative coefficient dominates over other contributions
$\propto \alpha^3$, this is more uncertain in QCD, where
$\log(1/\alpha_s)$ is much smaller. But
assuming that the unknown constant behind the log is roughly
of order 1, the off-diagonal susceptibilities
appear to be rather tiny in QCD (although they
would tend to become more important for larger $N$).
In fact, a most recent lattice study of nondiagonal
susceptibilities in $N_f=2$ QCD \cite{Gavai:2001ie} has found
only values consistent with zero, but
within statistical errors 
that are $\lesssim 10^{-6}$, whereas the natural order
of magnitude of (\ref{QCDtildechi}) is given by
${10\09\pi^3}[\alpha_s(2\pi T)]^3 
\sim 10^{-4}$ for $T\sim 3 T_c$.

The lattice calculations in Ref.~\cite{Gavai:2001ie} have been
performed with finite quark masses down to 
$m/T_c= 0.1$. However, this should not lead to any noticeable reduction,
because the first $m/T$ correction in (\ref{eff3m}) is
only quartic, leading to a correction factor
$\approx(1-0.06187 m^4/T^4)$ in the final result (\ref{QCDtildechi}), so 
the extreme smallness of the lattice
result of Ref.~\cite{Gavai:2001ie} remains a mystery for now.


\section{Conclusions}
To summarize, we have presented an analytical calculation
of the diagonal quark number susceptibility in hot QCD
within an approximately self-consis\-tent resummation of perturbation theory.
Our (non-perturbative) formulae include completely
the perturbative contributions of ${\mathcal  O}(g^2)$ and ${\mathcal O}(g^3)$. 
For
temperatures between $1.5$ to $5T_c$, our results show the same
general trend as seen on the lattice -- namely, a slow increase towards
the ideal gas results from below --, 
but with absolute values which are slightly,
but systematically, above the lattice data, by 5--10\%
in the case of $N_f=2$. 
This deviation is somewhat larger than that of our analogous
calculations of the entropy density 
\cite{Blaizot:1999ip,Blaizot:1999ap,Blaizot:2000fc}. However,
given that a continuum extrapolation of the lattice data
for quark susceptibilities
is still missing, it remains to be seen whether there are
sizable higher-order perturbative contributions 
not captured by our approach
or even important non-perturbative phenomena,
as speculated in Ref.~\cite{Gavai:2001ie}.

We have further computed the off-diagonal susceptibility
to lowest non-trivial order in perturbation theory, that is,
to order  $g^6\log(1/g)$. The result turns out to be
remarkably small, but not so small, however, to
explain the corresponding lattice result of Ref.~\cite{Gavai:2001ie},
which, surprisingly, is consistent with zero
with statistical errors $\lesssim 10^{-6}$.
This discrepancy certainly calls for more investigations
and more lattice data.

\section*{Acknowledgements}
Rajiv Gavai and Sourendu Gupta triggered our interest in this
problem. We would like to thank them for
discussions and for communication of their lattice results
prior to publication. We are also indebted to York Schr\"oder
for pointing out a sign error in eq.~(\ref{cc0pt})
of a previous version of this paper.

\end{document}